\documentclass[12pt,preprint]{aastex}

\newcommand{\eb}{\begin{equation}}
\newcommand{\ee}{\end{equation}}

\def\3dot#1{\stackrel{...}{#1}}

\newcommand{\msun}{$M_{\sun}$}

\newcommand{\masyr}{mas~yr$^{-1}$}

\slugcomment{Version Feb 10 2007}

\shorttitle{Astrometric orbits}
\shortauthors{Goldin \& Makarov}

\begin{document}

\title{Astrometric Orbits for Hipparcos Stochastic Binaries}

\author{A. Goldin \altaffilmark{1} \& V.V. Makarov \altaffilmark{2}}
\affil{$^1$ Citadel Investment Group, 131 South Dearborn st., Chicago, IL 60603}
\email{alexey.goldin@gmail.com}
\affil{$^2$Michelson Science Center, Caltech, 770 S. Wilson Ave.,
MS 100-22, Pasadena, CA 91125}
\email{vvm@caltech.edu}

\begin{abstract}
Taking advantage of an improved genetic optimization algorithm for fitting unconstrained
Kepler orbits to the Hipparcos intermediate astrometric data, we obtain additional
orbital solutions for 81 Hipparcos stars with previous stochastic (failed) solutions. 
The sample includes astrophysically interesting objects, including the old disk wide binary
HIP 754, the nearby AGB star HIP 34922 (L2 Pup), and the nearby M2 dwarf HIP 5496
(GJ 54, at 8 pc from the Sun) which has a resolved M dwarf companion.
\end{abstract}


\section{Introduction}
A genetic optimization-based algorithm of unconstrained orbital solutions on Hipparcos
Intermediate Astrometry Data \citep[HIAD,][]{esa} was developed and applied to 1561  stars from the annex of stochastic solutions, resulting in 65 orbiting binaries
at the 99\% confidence level \citep{goma}. Most of the 65 objects had not been known
as binaries, although at least two stars on the list of "new" binaries turned out to be parts of
a long-term spectroscopic monitoring carried out by other researches (R. Griffin,
J. Sperauskas, priv. comm.). Since our first publication about orbiting stochastic stars,
we have realized that a simple improvement of the algorithm can be effected if
the intermediate astrometric data produced by the two data analysis consortia (NDAC
and FAST) are combined prior to running the optimization program, as opposed to our
initial decision to run the program separately for the NDAC and FAST data points and
to compare the final results. Typically, each great circle measurement produced
a pair of abscissa data points, one derived by FAST, and the other by NDAC. Our initial disinclination 
to average the data derived from
the same observations was justified by the strong correlation which is underestimated
in the catalog. At the same time, some of the astrometric abscissae derived from the same observations
differ significantly between the two consortia, betraying unknown model or systematic
errors. These errors can be diminished by simple averaging, but the estimated precision
can only slightly improve because of the correlated noise. The safest and most conservative approach is to compute the mean values of the common data points,
but to adopt the larger of the two formal errors as the expected standard deviation of
the result, effectively assuming a correlation of 1. Thus,
our choice of methods and algorithms is driven by the intent to minimize the number of false
positives (which can not be precluded completely with these noisy data) and to produce confidence
estimates as realistic as possible.

\section{Uncertainties and bias}

Our results are affected by stochastic uncertainties, as well as by considerable bias, especially
in the important parameters $a_0$ (apparent orbit size) and $e$ (eccentricity). Therefore, it is of
crucial importance not only to compute an orbital fit to a set of data points, but also to
evaluate the probability distributions for the derived parameters. The one-dimensional nature
of Hipparcos measurements, the strong irregularity of observing times and position angles, and
the presence of unmodeled disturbances, such as visual companions, all combine to often produce
oddly shaped, asymmetric distribution functions. In order to compute realistic precision and bias
estimates, we use the parametric bootstrap method \citep{efr}.

Our general optimization method is described in \citep{goma}. First, we minimize
$\chi^2$ on abscissae residuals given observed transit times $t_i$ and transit times error
estimates $\sigma _i$. The best fit yields estimated times $t_{ei} $
such that $t ^j_i = t^j _{ei} + \sigma _i \epsilon^j _i$, where
$\epsilon ^j _i$ are normally distributed normal variables with
variance $1$ and $j$ is the number of realization. With sufficiently
large number of realizations, we build a histogram for each estimated fitting parameter.
The expected error of the estimate can be derived from the width of the histogram. 
In case of zero bias and nearly symmetric distribution, the mode of the distribution, its mean and median
are close to value derived from the original data. 

As an example, we consider the eccentricity fits produced by our unconstrained optimization
algorithm for a subset of 231 stars identified as orbital 
solutions in the main Hipparcos catalog \citep{esa}. Obviously, for high values of eccentricity 
the bias on the positive side is limited by the hard limit of 0.99 imposed by our
program. But if we consider only solutions with small enough eccentricity (less then 0.5), we can 
clearly see that only one solution out of 90 has a negative estimated bias. Only 3 solutions out of 
these 90 with small eccentricity have a negative estimated bias in semimajor axis $a_0$.

Due to the nonlinear mapping of $t_i$ to the orbital parameters, the
distributions of the mass function and $a_0$ may have extremely long
tails. Let $V_{est}$ be the estimated parameter value from the original data, and $V^j$ -- the estimate from a bootstrap realization $j$. To avoid the unwanted influence of occasional extreme values of $V$ on
our error and bias estimate, we define bias via the median, viz., $B(V) = V_{est} -
\hbox{median}(V)$, and $1\sigma$-error bars (which can be asymmetric) as
$\sigma_{-}(V)=\hbox{median}(V)-q_{0.16}(V)$ and
$\sigma_{+}(V)=q_{0.84}(V)-\hbox{median}(V)$. where $q_{g}(V)$ is the $g$-quantile of the
distribution of $V^j$.

\begin{figure}[htbp]
  \centering
  \includegraphics[angle=0,width=0.85\textwidth]{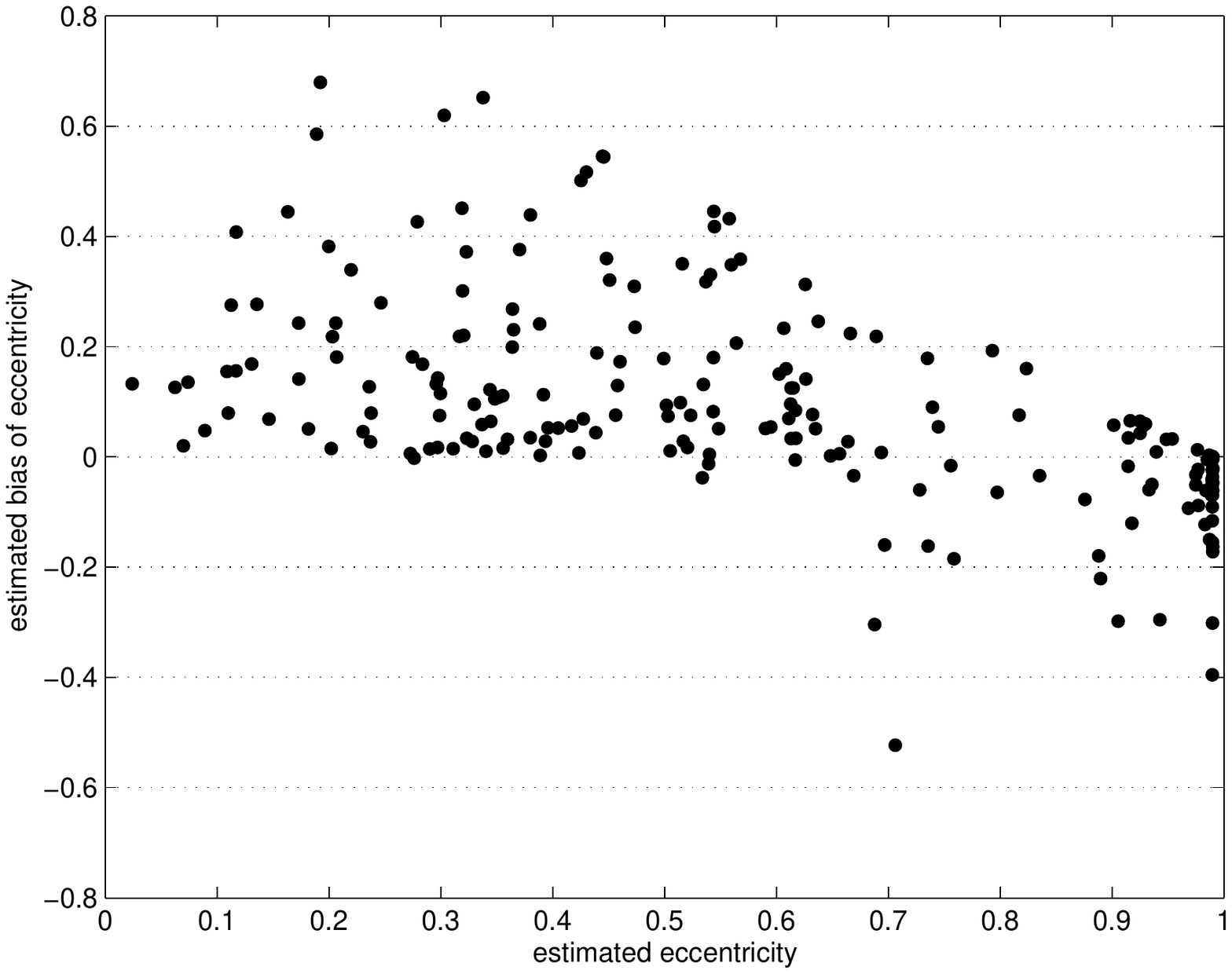}
  \caption{Orbital eccentricity and bias of eccentricity estimated by the parametric bootstrap
method for 231 binary stars with orbital solutions in the Hipparcos catalog.}
  \label{ebias.fig}
\end{figure}

For sufficiently small errors and sufficiently large number of data points, it
is possible to estimate the unbiased value of $V_{unbiased} = V_{est} -
B(V)$. We do not attempt this correction for the stochastic solutions considered in this work. 
For the 231 stars with catalogued Hipparcos orbital solutions, one can see from Fig.~\ref{ebias.fig}
that the bias of $e$ is
consistently positive if $e_{est}$ is not too large. Note that our eccentricity estimation
is restricted to the interval $[0,\,0.99]$, but the bias estimation is restricted only at the upper limit.
It is likely
that this bias is responsible for the abundance of solution with eccentricities close to the
cutoff value of $0.99$.  

\section{Orbits}
An automated run of the modified optimization algorithm on all 1561 stochastic stars in the
Hipparcos catalog resulted in about 200 new solutions with estimated confidence levels greater
than 99\%. The meaning of this confidence estimation is that the HIAD data for these stars
is perturbed in such a way that the apparent positions are consistent with the specified
orbital model, with the probability of a random occurrence of such data for a single, unperturbed
star less than 1\%. This is not to say that the given solutions are accurate to 99\%, or even
that the stars are 99\%-probability binaries. Indeed, we find that roughly one half of all
automated solutions from the stochastic set have eccentricities equal to $0.99$, and large orbital
sizes. This unexpected set of failed fits indicates that undetected orbital motion in our somewhat idealized model only accounts for a fraction of stochastic solutions. \citet{fama} demonstrated
that a number of stochastic stars are visual, resolved binaries with negligible
orbital motion, where the secondary companion was not listed in the Hipparcos input catalog,
or given at an incorrect position. Using correct initial assumptions about the relative position
of components in binary and multiple systems can yield regular, high-quality astrometric solutions
by a straightforward least-squares adjustment. But the model of fixed visual companions is another
simplification, bound to meet limited success with the real objects presenting a host
of complicating features. For example, a visual companion can have a detectable motion around
the primary, resulting in an effectively different proper motion; the inner pair in a hierarchical
triple system can be orbiting; one or both components of an orbiting pair can be variable stars 
evoking the variability-imposed motion (VIM) effect on the observed photocenter \citep{pour}.
For red stars, the data itself may need recalibration, since the assumed color was in many cases
biased \citep{pla}. In principle, increasingly sophisticated algorithms can be created to
deal with such cases; in practice though, the optimization problem becomes cumbersome with the
available data because we are quickly running out of data points with each additional nonlinear
fitting parameter.

At this point, we simply rejected all the spurious-looking solutions with extreme eccentricities.
The remaining $81$ fits are specified in Table~\ref{big.tab}. Orbital fits for the stars
published in our preceding paper \citep{goma} are also omitted, even if they differ for some
stars. The estimated astrometric
parameters, period $P$ in days, eccentricity $e$, periastron time $T_0$ in days, longitude of the ascending node $\omega$ in degrees, position angle of the node $\Omega$ in degrees, and inclination $i$ (zero for face-on orbits) in degrees are specified in the table. These data are based on 1000 genetic 
optimization trials for each object. Confidence limits for each fitting parameter are computed at several confidence levels
from the set of 1000 simulations, but only the 84th and 16th percentiles are used to
compute the robust uncertainty limits in
Table~\ref{big.tab}, which correspond to the $\pm 1\sigma$ interval of $N(m,\sigma)$. These intervals
provide a minimalistic estimation of the parameter uncertainties, but carry no information about
the shape of the distribution, which can be strongly non-Gaussian, as discussed in the previous
paragraph. Our solutions are perfectly
unconstrained, and are not based on any prior information except the HIAD.

A literature search for the stars listed in Table~1 reveals that three objects have been
known as spectroscopic binaries from precision radial velocity observations. Two of them
have accurate spectroscopic orbits published, viz., {\bf HIP 34164 = G 108-53} \citep{lat}
and {\bf HIP 67480 = e Boo} \citep{gri}. Both stars were also re-analyzed by \citet{jan},
whose computations were also based on the HIAD, but the essential nonlinear parameters ($P$,
$e$ and $\omega$) were fixed at their spectroscopic values. The cited spectroscopic
data, our unconstrained fits and constrained astrometric solutions from \citep{jan} are
specified in Table~\ref{known.tab} for comparison. Encouragingly, our periods come out fairly
close (within 1-2 $\sigma$) to the more accurate spectroscopic estimates, and the eccentricities
are even closer to the mark for both stars. The unconstrained estimates of $\omega$ probably
include the uncertainty of $180\degr$ in both cases. Therefore, we do not see any reasons to
be concerned with our method and results from the comparison with the spectroscopic data.
The constrained solutions for the apparent orbit size $a_0$ and inclination $i$ from \citep{jan}
are much stronger than ours, because fixing the troublesome nonlinear parameters (especially,
$e$) simplifies the multidimensional $\chi^2$ surface and removes many of the valley-like local
minima. For HIP 34164, our estimate of $i$ is comfortably close to the constrained parameter.
The corrected mass function, $M_2^3/(M_1+M_2)^2$, is fairly high at $0.095$ $M_{\sun}$ for
this high proper motion, moderately metal-poor ([Fe/H]$=-0.68$) system.

For the star HIP 67480 (e Boo), our unconstrained fit and the constrained solution from \citep{jan}
produce similar orbit sizes, but significantly different inclination angles. The former value
of $i$ is brought into agreement with the small radial velocity amplitude reported by \citet{gri}
for this K4III star and the spectroscopic $a_1\sin i\approx 14.1$ Gm. Our unconstrained solution
would imply a companion of moderate mass in the absence of a spectroscopic data, whereas we are probably
dealing with an orbit seen almost face-on.

Finally, the F7V star {\bf HIP 110785} = HD 212754 has been listed by \citet{gris} as a single-lined
spectroscopic binary, but without orbital solution. This star has been drawing considerable attention
of observers since it was identified as a likely member of the Hyades moving group. Our solution yields
an almost edge-on orbit of size $a_0\approx 0.43$ AU. From this, and the period $P=878$ d, we
conclude that the mass ration $M_2/M_1$ may be fairly small ($\simeq 1/5$).

The Geneva-Copenhagen (GC) spectroscopic survey of Hipparcos solar-type stars
\citep{nord} provides an independent verification of our results. Most of the stars in this
survey obtained only a few radial velocity (RV) measurements, which was insufficient to
derive an orbital fit, but sufficient to establish the fact of spectroscopic binarity. For stars with
more than one measurement the probability is specified that the RV has not changed significantly 
relative to the estimated measurement error. These probability estimates are useful to
make sure that most of our orbital solutions indeed correspond to binary systems, although the
quality of our solutions can not be verified. We find 15 stars in common between Table 1 and GC
with estimated probabilities of constant RV,
viz., HIP 754, 1768, 3645, 12726, 17482, 32307, 49638, 56447, 64790, 84696, 101430, 110785, 113177,  and 113699. All these probabilities are zero or very small in GC. We conclude that the incidence of
spectroscopic binarity among the new solutions is probably high.

The stars HIP 3645, 43067, 49638, 50567, 84696, 109095 are listed in \citep{maka} as astrometric
binaries whose long-term proper motions from the Tycho-2 catalog are statistically different from the short-term Hipparcos proper motion. Such discrepancies are usually caused by significant long-period
orbital motion. The fitted periods for these stars are indeed at the upper end of the range in
Table~1, exceeding 3-4 years. Admittedly, this is not a clean check on the quality and reliability
of our solutions, because the stochastic proper motions in Hipparcos can be perturbed more than
their formal errors manifest.

The data in Table 1 indicate that our method is especially sensitive for orbital periods between
1 and 4 years, with relatively few shorter or longer periods detected. The smallest orbital
size appears to be 5 mas. It is difficult to derive robust orbital solutions for smaller 
apparent orbits even for the brightest stars, where the typical single measurement precision
is comparable to or less than 5 mas. This is an interesting fact pertaining to the scarcity of new low-mass
companions detected from the HIAD, and the complete lack of planet signatures. Rather than a hint
of lower than expected precision of the data, we consider it to be a still poorly understood
feature of the given nonlinear optimization problem, which has been empirically found to have
a worse condition than what could be naively expected from the number of data points and the number
of fitting parameters \citep[see also][]{pour}.

\begin{figure}[htbp]
  \centering
  \includegraphics[angle=0,width=0.85\textwidth]{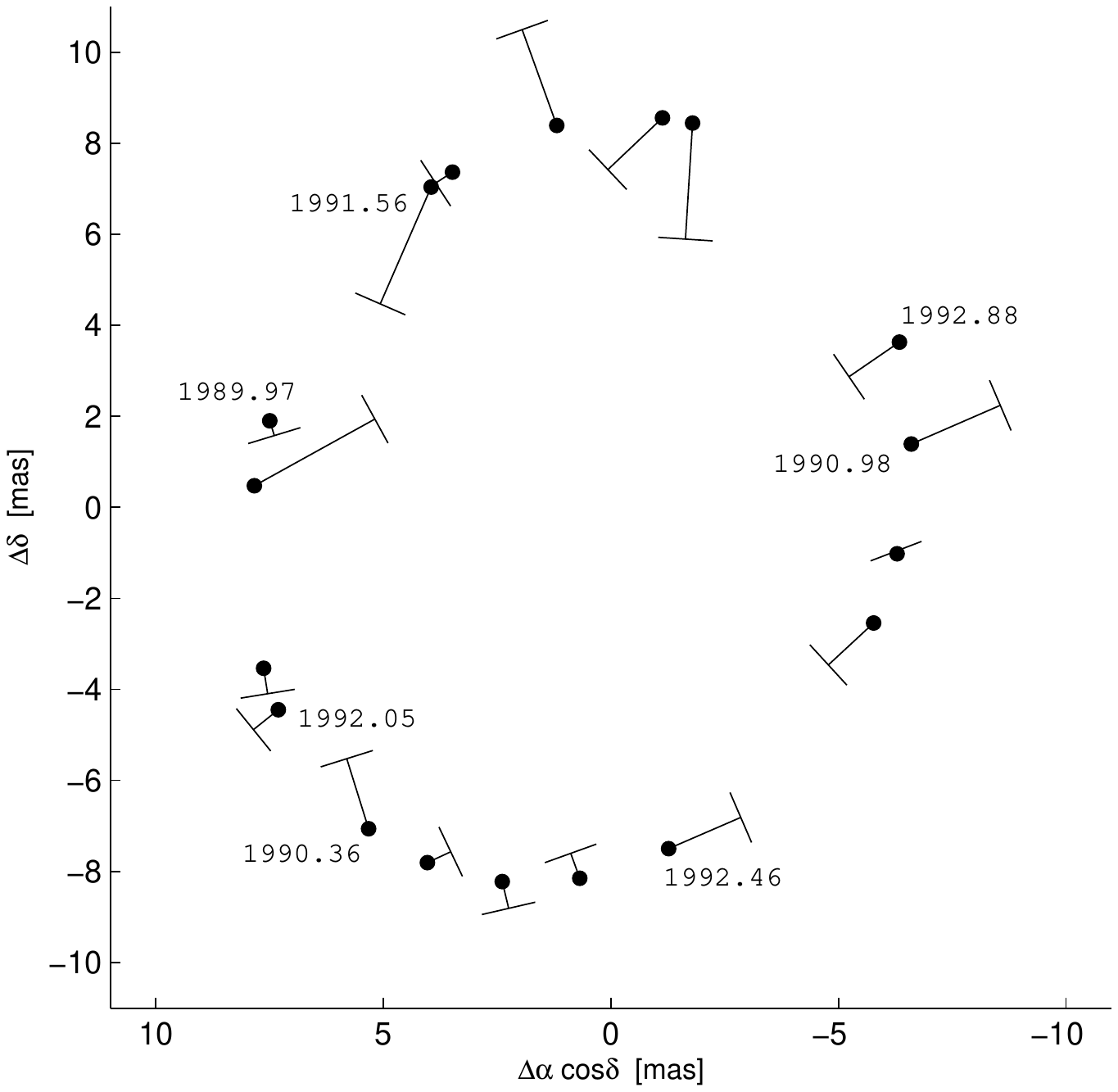}
  \caption{The apparent orbit of the photocenter of the new binary HIP 113699. The observed abscissa residuals
are shown with straight segments connecting the estimated orbital positions (indicated with
black circles) and the actual observations. The mean epoch positions of the Hipparcos
grid of slits is marked with short straight segments. The $\chi^2$-adjusted parameters (slightly
different from those specified in Table 1) are: $P=667$ d, $e=0.1$, $T_0=216$ d, $i=36.4\degr$,
$a_0=8.7$ mas, $\omega=97\degr$.}
  \label{orb.fig}
\end{figure}

The orbital fits for a few stars in Table~1 are so definite and robust, that one may wonder why
they were not resolved by the Hipparcos data analysis team and placed in the annex of orbital
solutions. Fig.~\ref{orb.fig} shows the reconstructed orbit and the averaged observations for
one of these stars, HIP 113699. Our fit yields an orbital period of nearly 2 years and an apparent
motion of $a_0=9.5$ mas for this poorly investigated G0 dwarf. The star was part of the systematic
attempt to resolve Hipparcos stochastic binaries with speckle interferometry by \citet{mas},
which was unsuccessful. The estimated orbit size in combination with the parallax (23 mas) indicates
that the companion may be much fainter than the primary. The estimated inclination suggests that
the star should be easily detectable as SB1 in RV measurements.

\section{Objects of note}

\begin{figure}[htbp]
  \centering
  \includegraphics[angle=0,width=0.85\textwidth]{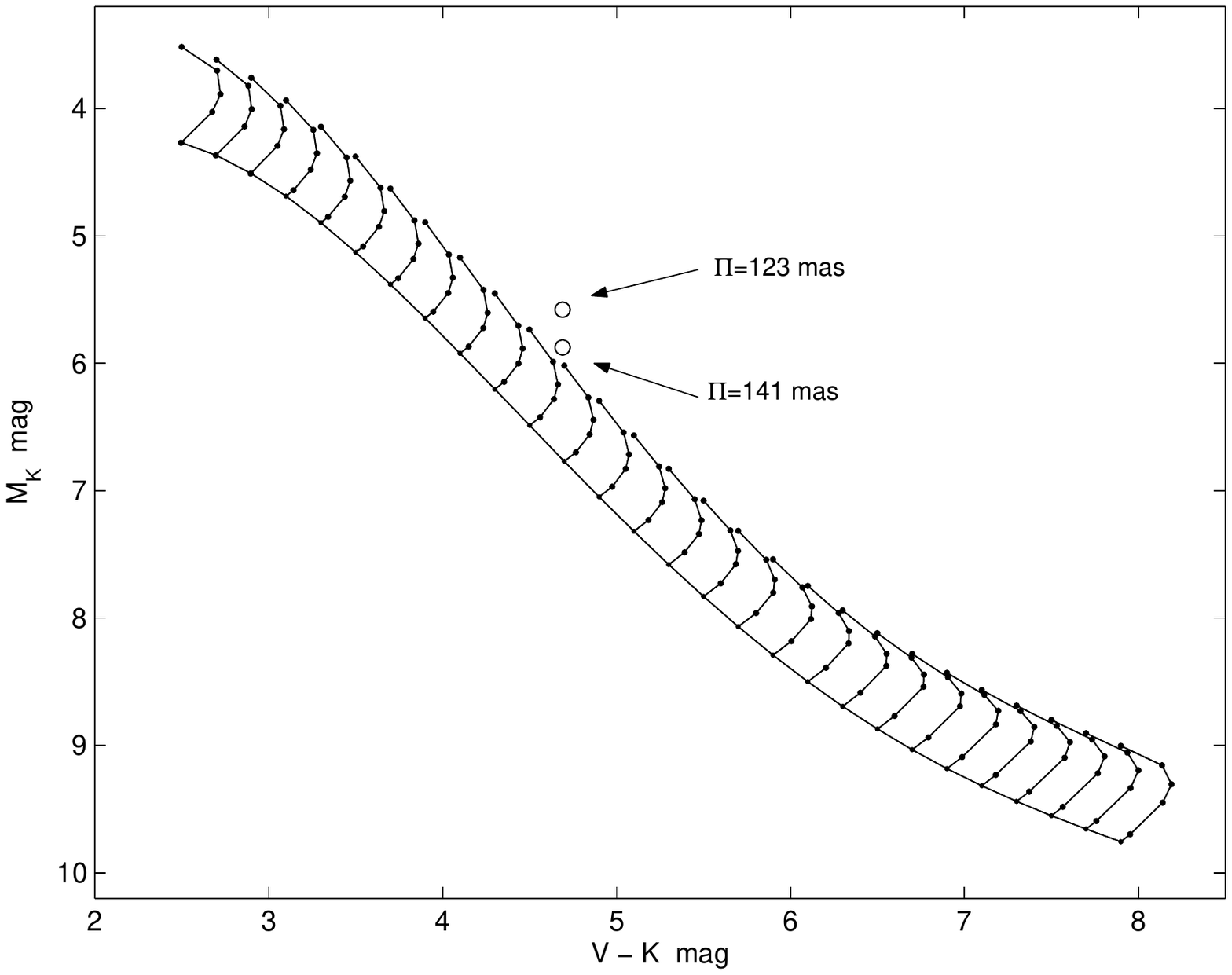}
  \caption{The empirical main sequence of field dwarfs from \citep{hen04} (lower curve)
and the location of binaries with joint photometric magnitudes shown with short
curves and dots for $\Delta V=$ 0 (upper end points), 1, 2, 3, 6 and $+\infty$ mag. The loci of
the star GJ 54 are indicated with open circles for two parallaxes, 123 and 141 mas.}
  \label{bin.fig}
\end{figure}

{\bf HIP 5496 = GJ 54} is a nearby M dwarf at a distance of $\approx 8$ pc from the Sun.
The stochastic Hipparcos parallax $123\pm 8$ mas is somewhat uncertain, but a recent
ground-based observation puts the star confidently within 10 pc \citep{hen}. \citet{gol}
resolved a visual companion to this star using {\it HST} NICMOS observations and found
a magnitude difference of $\approx 1$ mag between the components. Our solution for this star
is relatively robust, with a period of $427(+9-8)$ d and apparent semimajor axis $a_0=
19.0(+7.4,-2.1)$ mas. The small orbital size for this nearby star seems to
suggest a low-mass companion, perhaps a brown dwarf, raising doubts that the orbiting
companion and the visual are the same one. However, the provisional data from \citep{hen}
confirm our orbital solution, estimating the period at $\approx 1.1$ year and the apparent
orbit of the photocenter at 16 mas. Obviously, the companion is bright enough (magnitude difference of 
1.04 in $V$) to pull the photocenter closer to the center of mass, making the observed ellipse
much smaller than the true one. A simple calculation shows that for a plausible total
mass of $0.5$ $M_{\sun}$, the full orbit is $a=0.88$ AU, and since the observed photocenter
orbit is $0.15$ AU, the primary's orbit is $0.40$ AU, and the mass ratio $q=M_2/M_1=0.83$.
Thus, the Hipparcos measurements and our orbital solution are consistent with the known properties
of the resolved companion. What remains puzzling is the overall brightness of the system. According
to \citep{hen}, the magnitudes are $V=9.82$, $I=7.32$, $J=6.00$, and $K=5.13$ mag. This compares
well with the accurate photometry from \citep{koen}, $V=9.812\pm0.008$ and $I=7.357\pm0.009$ mag.
Fig.~\ref{bin.fig} depicts the well-defined empirical main sequence of nearby dwarfs in the
$M_K$ vs. $V-K$ axes from \citep{hen04} (lower curve). The short curled curves with beads on them
indicate the joint photometry of unresolved main-sequence binaries on a grid of primary
absolute magnitudes, for magnitude differences between the companions of $\Delta V=+\infty$, 6, 3,
2, 1, and 0 mag. The latter value corresponds to the case of two equally bright companions (twins).
With the trigonometric parallax determined by Hipparcos, the system is much brighter in $K$ than
the upper envelope of possible magnitudes for binary stars. The parallax measured by \citet{hen}
($141.2\pm3.4$ mas) brings it closer to the range of normal magnitudes, but interestingly,
our updated astrometric solution for parallax is close to the original Hipparcos value. \citet{hen} contend
that this system represents almost the worst case, when the orbital period is close to one year,
and it becomes difficult to distinguish between the parallactic and orbital ellipses. The covariance
analysis technique on linearized condition equations described in \citep{goma} is useful to
estimate the degree of possible perturbation of the parallax solution. We compute a significant
correlation coefficient of $-0.24$ for the pair of parameters $a_0$ and $\Pi$. It can be suspected
that the trigonometric parallax is underestimated; therefore, the orbit size may be slightly overestimated. 
We also find, surprisingly, large positive correlations between $\Pi$ and the angles $\omega$ and
$\Omega$ of 0.57 and 0.61, respectively. It follows, that the actual difficulty with this star is
that the parallax determination is strongly coupled with the orientation of the orbit, which is
poorly constrained by our data.

{\bf HIP 34922 = L$_2$ Pup} is probably the nearest asymptotic giant branch (AGB) star to
the Sun. AGB stars are in the terminal stages of stellar evolution characterized by massive
loss of mass. By virtue of its large-scale, stable variability of brightness, the star has been
classified as semiregular pulsating variable. Our unconstrained fit for this star produced a robust
solution with a period of $141(+2,-1)$ days, which is identical to the variability period of 140.6
d. This prompted us to investigate if the star could be a Variability-Imposed-Motion (VIM) binary,
whose photocenters move along the line connecting the components in phase with the variability,
because of the shift of the center of light. A general VIM fit produced a meagre reduction in the
residual $\chi^2$, ruling out a simple VIM model. Another possibility of a spurious astrometric fit
is predicated by the incorrect $V-I$ color indeces that had been assumed for some of the red stars
in the data reduction. This could generate significant errors in color-dependent calibration
terms, as explained in detail in \citep{pla}. For this star, however, the assumed color ($V-I=4.18$)
was not too different from the color derived from subsequent photometric observations ($V-I=3.51$),
and the calibration error is not expected to exceed 1 mas, which is much smaller than the
fitted orbit ($9.5\pm[+4.1-1.3]$). Thus, L$_2$ Pup is quite likely a real orbiting binary. \citet{jur}
obtained and studied mid-infrared images of L$_2$ Pup and found an extended asymmetric
morphology of the shell. The asymmetry was later confirmed from the interferometric observations by
\citet{ire}. \citet{leb} detected periodic variation of radial velocity, but unfortunately, the small
number of data points does not permit to estimate the phase shift between the light curve and the
RV curve. We propose that the stable period of the variability curve, the recent dimming of the object
and other peculiar properties of the object can be explained by a bright companion orbiting the
primary AGB star and getting periodically obscured by its thick dust shell. A more detailed discussion
of this system will be published elsewhere.

{\bf HIP 754 = HD 471} has been netted in earlier searches for field subdwarfs because of its
considerable proper motion \citep[e.g.,][]{san}. However, the star is moderately metal poor at
[Fe/H]$=-0.16$ \citep{nord}, has a small Galactic eccentricity of 0.11, and its maximum excursion
from the Galactic plane is limited to 258 pc \citep{all}. Thus, it is not a subdwarf, but Allen et
al. argue that it is one of the oldest known wide binaries with a photometrically estimated age
of 11.6 Gyr. The wide companion is separated by a little less than $30\arcsec$. This is definitely a
physical pair, with proper motions from the extensive Nomad catalog \citep{zac} $(\mu_{\alpha *},
\mu_\delta)=(174.9,-146.7)\pm(1.8,1.3)$ \masyr~ for HIP 754 A and $(171.0,-147.5)\pm(0.6,0.9)$ \masyr~ for HIP 754 B. The wide companion has a separation of at least 1200 AU, and a period $\ga 34000$ yr.
Dynamical evolution and the origin of old wide binaries are interesting issues motivating the long-term
effort to increase the sample of such systems and to collect more observations. Given the considerable
attention to this star in the literature, it is surprising that binarity of the primary in this pair
has remained unnoticed. Our solution suggests a fairly large apparent orbit of $9.8\pm(+3.9,-1.5)$
mas and a nearly edge-on position, explaining the large scatter of radial velocities observed in \citep{nord}. The corrected parallax is $24.8\pm 1.0$ mas. Assuming a total mass of 1.5 \msun, the mass of the primary
is roughly 1.1 \msun~ and the secondary 0.4 \msun. \citet{tok} argue that new stars do not form in
short-period binaries, and we mostly owe the presence of spectroscopic binaries with periods less than 10 d
to the dynamical interaction between the inner pair and the remote tertiary companion in
hierarchical multiple systems, which takes giga-years to shrink the inner orbit, and perhaps, to
eventually merge the close pair. The HIP 754 system, if it really is very old, appears to be an
example to the contrary, when the tertiary failed for some reasons to shrink the close inner binary.

{\it Facility:} \facility{HIPPARCOS}

\acknowledgments
The research described in this paper was in part carried out at the Jet Propulsion 
Laboratory, California Institute of Technology, under a contract with the National 
Aeronautics and Space Administration. This research has made use of the SIMBAD database,
operated at CDS, Strasbourg, France.

\begin{deluxetable}{r | rr |  rr |  rr |  rr | rr | rr | rr}
\tabletypesize{\scriptsize}
\rotate
\tablecaption{Orbital solutions. \label{big.tab}}
\tablewidth{0pt}
\startdata

\multicolumn{1}{c}{ HIP}&  \multicolumn{2}{c}{$P\pm\sigma$}  &  \multicolumn{2}{c}{$T_0\pm\sigma$} & \multicolumn{2}{c}{$a_0\pm\sigma$}   &  \multicolumn{2}{c}{$e\pm\sigma$}  &  \multicolumn{2}{c}{$i\pm\sigma$}  &  \multicolumn{2}{c}{$\omega\pm\sigma$}  &  \multicolumn{2}{c}{$\Omega\pm\sigma$}\\
\multicolumn{1}{c}{}& \multicolumn{2}{c}{d}  &  \multicolumn{2}{c}{d}  &  \multicolumn{2}{c}{mas}  &  \multicolumn{2}{c}{}  &  \multicolumn{2}{c}{$\degr$}  &  \multicolumn{2}{c}{$\degr$}  &  \multicolumn{2}{c}{$\degr$}  \\
\hline
 & & & & & & & & & & & & & & \\
754	 & 483	 & +12-13	 & 216	 & +66-82	 &	9.8	& +3.9-1.5	&0.22	& +0.41-0.20 & 74	 & +9-11 & 112	 & +46-45 & 169	 & +19-12\\
1768	 & 1150	 & +46-34	 & 876	 & +141-402	 &	40.0	& +2.2-2.1	&0.07	& +0.08-0.07 & 47	 & +4-4	 & 117	 & +52-75 & 93	 & +177-8\\
3645	 & 1282	 & +580-132	 & 598	 & +643-433	 &	21.2	& +5.6-1.6	&0.11	& +0.33-0.14 & 86	 & +5-5	 & 102	 & +68-65 & 129	 & +175-4\\
3750	 & 1748	 & +764-399	 & 758	 & +634-576	 &	17.2	& +13.3-2.9	&0.77	& +0.22-0.21 & 81	 & +5-6	 & 33	 & +23-14 & 80	 & +5-7\\
4365	 & 920	 & +60-59	 & 553	 & +305-459	 &	19.0	& +3.7-2.7	&0.15	& +0.27-0.16 & 121	 & +9-8	 & 101	 & +72-70 & 132	 & +175-20\\
5496	 & 427	 & +9-8	         & 182	 & +65-48	 &	19.0	& +7.4-2.1	&0.41	& +0.33-0.21 & 115	 & +8-8	 & 141	 & +32-108 & 75	 & +180-14\\
6542	 & 1803	 & +452-279	 & 432	 & +431-213	 &	22.8	& +7.7-3.1	&0.69	& +0.16-0.18 & 52	 & +8-7	 & 106	 & +21-16 & 138	 & +11-11\\
8486	 & 2471	 & +1782-527	 & 964	 & +1710-387	 &	53.2	& +16.7-4.3	&0.45	& +0.25-0.21 & 64	 & +2-2	 & 65	 & +166-22 & 333 & +5-176\\
11925	 & 928	 & +285-31	 & 799	 & +92-635	 &	13.4	& +12.7-2.6	&0.56	& +0.24-0.18 & 93	 & +5-4	 & 106	 & +160-16 & 351 & +7-348\\
12225	 & 1098	 & +60-66	 & 932	 & +115-751	 &	23.1	& +1.6-1.0	&0.16	& +0.16-0.11 & 117	 & +3-3	 & 83	 & +23-34 & 56	 & +4-4\\
12726	 & 536	 & +12-12	 & 301	 & +51-47	 &	11.7	& +3.8-1.6	&0.61	& +0.19-0.13 & 145	 & +10-12 & 109	 & +50-51 & 86	 & +93-47\\
12894	 & 841	 & +30-30	 & 490	 & +134-121	 &	10.1	& +1.4-1.2	&0.28	& +0.19-0.15 & 123	 & +8-7	 & 114	 & +154-45 & 285 & +15-172\\
16192	 & 695	 & +24-25	 & 404	 & +185-179	 &	13.3	& +3.5-2.0	&0.08	& +0.30-0.17 & 56	 & +8-9	 & 82	 & +96-49 & 140	 & +174-22\\
17022	 & 1093	 & +101-62	 & 350	 & +237-166	 &	12.8	& +4.9-1.7	&0.49	& +0.31-0.18 & 66	 & +6-7	 & 140	 & +135-121 & 252 & +19-175\\
17482	 & 699	 & +26-28	 & 186	 & +327-128	 &	13.2	& +3.2-1.8	&0.27	& +0.24-0.19 & 132	 & +10-9 & 100	 & +151-41 & 205 & +28-158\\
19673	 & 744	 & +60-45	 & 564	 & +140-497	 &	6.5	& +2.5-1.2	&0.44	& +0.24-0.21 & 41	 & +15-12 & 109	 & +150-44 & 222 & +47-158\\
19832	 & 664	 & +26-26	 & 257	 & +287-163	 &	18.2	& +2.6-1.8	&0.05	& +0.18-0.12 & 49	 & +10-10 & 74	 & +100-46 & 174 & +170-155\\
21386	 & 1030	 & +22-24	 & 893	 & +83-136	 &	19.9	& +1.1-1.0	&0.20	& +0.09-0.08 & 99	 & +2-2	 & 48	 & +150-29 & 203 & +4-178\\
21965	 & 709	 & +26-23	 & 471	 & +105-114	 &	9.5	& +2.6-1.3	&0.29	& +0.30-0.17 & 127	 & +9-10 & 107	 & +41-40 & 161	 & +16-19\\
30223	 & 796	 & +47-49	 & 266	 & +194-132	 &	8.4	& +3.7-1.4	&0.24	& +0.36-0.23 & 51	 & +12-12 & 88	 & +72-59 & 153	 & +172-39\\
30342	 & 452	 & +13-16	 & 311	 & +84-232	 &	5.9	& +1.6-0.9	&0.20	& +0.35-0.19 & 116	 & +12-10 & 109	 & +53-53 & 53	 & +166-16\\
32307	 & 889	 & +53-56	 & 468	 & +240-287	 &	9.2	& +1.6-1.0	&0.15	& +0.25-0.16 & 43	 & +12-12 & 105	 & +69-65 & 166	 & +76-144\\
34036	 & 606	 & +89-70	 & 334	 & +158-250	 &	12.9	& +17.6-3.9	&0.33	& +0.41-0.25 & 71	 & +13-12 & 110	 & +62-64 & 121	 & +165-21\\
34164	 & 643	 & +91-51	 & 426	 & +142-304	 &	9.3	& +19.8-2.3	&0.27	& +0.42-0.31 & 104	 & +15-11 & 111	 & +65-53 & 103	 & +116-76\\
34922	 & 141	 & +2-1	 	 & 118	 & +17-106	 &	9.5	& +4.1-1.3	&0.52	& +0.30-0.21 & 118	 & +8-9    & 131 & +162-38 & 239 & +15-169\\
36399	 & 1001	 & +25-27	 & 782	 & +192-760	 &	17.1	& +5.8-2.4	&0.62	& +0.17-0.13 & 139	 & +11-13 & 124	 & +159-32 & 249 & +33-160\\
37272	 & 798	 & +49-32	 & 446	 & +164-104	 &	26.5	& +28.9-6.6	&0.72	& +0.17-0.29 & 77	 & +8-13 & 93	 & +81-71 & 134	 & +182-18\\
38018	 & 547	 & +8-9	         & 476	 & +29-33	 &	17.7	& +1.7-1.1	&0.52	& +0.11-0.09 & 45	 & +7-7	 & 39	 & +167-19 & 258 & +12-171\\
38146	 & 889	 & +17-22	 & 545	 & +58-81	 &	11.7	& +1.5-1.1	&0.47	& +0.15-0.14 & 42	 & +10-11 & 153	 & +49-134 & 186 & +140-164\\
38910	 & 2772	 & +2190-807	 & 1549	 & +1824-856	 &	47.6	& +20.4-7.8	&0.40	& +0.30-0.19 & 49	 & +7-9	 & 132	 & +45-61 & 176	 & +30-152\\
38980	 & 927	 & +141-66	 & 557	 & +204-314	 &	13.1	& +5.9-1.9	&0.77	& +0.28-0.27 & 77	 & +6-7	 & 86	 & +156-33 & 295 & +9-174\\
39681	 & 1684	 & +962-296	 & 916	 & +619-699	 &	19.5	& +10.6-2.3	&0.54	& +0.26-0.27 & 68	 & +7-7	 & 88	 & +29-27 & 58	 & +10-7\\
39903	 & 925	 & +12-12	 & 164	 & +65-65	 &	27.3	& +0.8-0.8	&0.08	& +0.05-0.04 & 30	 & +4-5	 & 56	 & +158-28 & 340 & +12-180\\
42728	 & 403	 & +24-16	 & 247	 & +113-196	 &	7.9	& +9-2.8	&0.31	& +0.36-0.34 & 85	 & +12-14 & 101	 & +72-62 & 134	 & +170-18\\
42916	 & 815	 & +13-14	 & 331	 & +44-49	 &	21.4	& +4.3-1.8	&0.68	& +0.10-0.07 & 149	 & +9-12 & 114	 & +57-47 & 168	 & +48-97\\
42984	 & 455	 & +15-15	 & 124	 & +245-80	 &	5.2	& +5.2-1.0	&0.27	& +0.44-0.24 & 133	 & +13-17 & 92	 & +67-50 & 122	 & +86-77\\
43067	 & 1592	 & +1263-349	 & 845	 & +748-579	 &	11.6	& +8.4-2.9	&0.40	& +0.26-0.26 & 58	 & +9-10 & 103	 & +32-34 & 91	 & +18-14\\
43352	 & 1191	 & +247-136	 & 444	 & +514-245	 &	9.0	& +2.0-1.3	&0.54	& +0.14-0.13 & 132	 & +12-9 & 106	 & +167-23 & 204 & +17-168\\
49638	 & 1546	 & +372-195	 & 786	 & +569-681	 &	22.3	& +12.2-3.2	&0.67	& +0.23-0.21 & 117	 & +6-8	 & 101	 & +172-13 & 274 & +9-173\\
50180	 & 1590	 & +1168-231	 & 897	 & +681-543	 &	32.3	& +14.3-3.0	&0.18	& +0.26-0.13 & 79	 & +4-5	 & 154	 & +136-83 & 193 & +5-177\\
50567	 & 1102	 & +180-73	 & 324	 & +470-191	 &	14.4	& +3.3-1.8	&0.40	& +0.23-0.18 & 40	 & +13-11 & 85	 & +62-48 & 66	 & +199-31\\
53256	 & 743	 & +31-30	 & 131	 & +179-71	 &	8.1	& +9.5-1.5	&0.42	& +0.38-0.22 & 48	 & +19-13 & 74	 & +41-40 & 49	 & +130-27\\
54424	 & 1255	 & +241-147	 & 805	 & +355-345	 &	9.5	& +2.8-1.4	&0.51	& +0.20-0.18 & 133	 & +12-10 & 109	 & +164-26 & 296 & +26-165\\
56447	 & 526	 & +24-27	 & 290	 & +168-233	 &	18.6	& +19.6-3.5	&0.60	& +0.34-0.31 & 65	 & +11-11 & 83	 & +43-26 & 140	 & +15-11\\
61100	 & 1366	 & +199-127	 & 171	 & +564-104	 &	35.2	& +4.8-3.2	&0.63	& +0.09-0.07 & 61	 & +5-5	 & 63	 & +12-11 & 175	 & +5-6\\
62512	 & 1574	 & +238-100	 & 217	 & +400-116	 &	43.9	& +3.1-1.4	&0.14	& +0.16-0.09 & 107	 & +2-2	 & 164	 & +11-114 & 30	 & +178-2\\
64648	 & 1267	 & +441-147	 & 801	 & +374-522	 &	8.8	& +8.5-1.7	&0.29	& +0.44-0.29 & 99	 & +8-6	 & 81	 & +49-34 & 131	 & +31-11\\
64790	 & 147	 & +1-1	         & 31	 & +23-16	 &	7.6	& +3.9-1.5	&0.38	& +0.28-0.25 & 58	 & +11-11 & 81	 & +27-35 & 74	 & +16-12\\
67480	 & 1167	 & +153-89	 & 763	 & +287-259	 &	9.6	& +6.0-1.8	&0.38	& +0.35-0.21 & 134	 & +13-15 & 111	 & +45-42 & 90	 & +87-30\\
76006	 & 572	 & +21-21	 & 314	 & +90-81	 &	6.0	& +15.2-1.2	&0.51	& +0.33-0.27 & 55	 & +17-13 & 105	 & +62-60 & 89	 & +119-41\\
78970	 & 919	 & +38-40	 & 792	 & +97-727	 &	11.0	& +1.2-1.0	&0.09	& +0.18-0.12 & 61	 & +6-7	 & 113	 & +66-79 & 55	 & +163-26\\
80884	 & 1021	 & +125-46	 & 176	 & +656-117	 &	13.7	& +2.3-1.4	&0.46	& +0.17-0.14 & 66	 & +6-7	 & 125	 & +166-23 & 293 & +8-176\\
84062	 & 730	 & +33-32	 & 110	 & +559-85	 &	10.4	& +2.0-1.3	&0.38	& +0.27-0.19 & 55	 & +9-10 & 51	 & +117-34 & 29	 & +185-13\\
84223	 & 2852	 & +2511-987	 & 1589	 & +2358-889	 &	24.4	& +14.7-5.8	&0.41	& +0.26-0.29 & 111	 & +10-9 & 103	 & +75-67 & 175	 & +170-25\\
84696	 & 1448	 & +880-242	 & 954	 & +423-791	 &	23.8	& +7.6-2.2	&0.22	& +0.38-0.24 & 120	 & +6-6	 & 128	 & +48-50 & 176	 & +179-171\\
84949	 & 1648	 & +582-239	 & 348	 & +612-199	 &	16.5	& +9.7-2.9	&0.67	& +0.18-0.14 & 64	 & +8-6	 & 73	 & +18-15 & 144	 & +7-9\\
85852	 & 950	 & +47-47	 & 588	 & +287-494	 &	7.27	& +1.1-0.9	&0.16	& +0.21-0.14 & 133	 & +12-10 & 111	 & +102-74 & 171 & +163-36\\
85954	 & 225	 & +520-3	 & 144	 & +411-37	 &	14.5	& +16.5-2.6	&0.42	& +0.48-0.23 & 65	 & +38-12 & 92	 & +59-48 & 152	 & +169-14\\
88848	 & 1715	 & +300-179	 & 267	 & +686-169	 &	37.4	& +6-3.3	&0.61	& +0.12-0.13 & 77	 & +3-2	 & 111	 & +174-11 & 251 & +4-178\\
90355	 & 302	 & +7-6	         & 182	 & +55-43	 &	9.7	& +2.5-1.4	&0.35	& +0.21-0.17 & 139	 & +12-13 & 86	 & +90-54 & 168	 & +57-67\\
93137	 & 864	 & +103-43	 & 494	 & +263-133	 &	11.6	& +3.0-1.6	&0.54	& +0.19-0.19 & 83	 & +5-6	 & 99	 & +26-19 & 30	 & +6-5\\
94347	 & 636	 & +16-15	 & 375	 & +72-59	 &	12.8	& +1.6-1.1	&0.30	& +0.15-0.14 & 145	 & +9-9	 & 68	 & +42-30 & 34	 & +144-21\\
94802	 & 1037	 & +38-30	 & 521	 & +191-254	 &	18.2	& +1.0-0.9	&0.09	& +0.09-0.06 & 132	 & +5-4	 & 90	 & +63-52 & 90	 & +172-7\\
95044	 & 901	 & +66-52	 & 176	 & +480-116	 &	16.1	& +4.3-2.2	&0.45	& +0.24-0.21 & 78	 & +5-5	 & 79	 & +163-27 & 247 & +7-176\\
97063	 & 1297	 & +311-99	 & 570	 & +526-213	 &	17.3	& +2.6-1.5	&0.38	& +0.17-0.14 & 33	 & +9-10 & 126	 & +160-42 & 222 & +73-170\\
97690	 & 1051	 & +72-61	 & 499	 & +384-341	 &	8.0	& +1.1-0.9	&0.08	& +0.22-0.12 & 61	 & +7-8	 & 105	 & +69-66 & 44	 & +173-17\\
98375	 & 284	 & +7-6	         & 203	 & +49-125	 &	5.9	& +6.0-1.2	&0.44	& +0.33-0.29 & 91	 & +8-7	 & 90	 & +160-35 & 194 & +13-174\\
101430	 & 1437	 & +270-145	 & 1068	 & +294-904	 &	15.5	& +4.5-1.9	&0.55	& +0.23-0.23 & 93	 & +4-4	 & 78	 & +23-19 & 178	 & +6-7\\
103287	 & 1295	 & +197-99	 & 888	 & +341-544	 &	28.4	& +3.2-1.7	&0.23	& +0.23-0.16 & 89	 & +4-4	 & 124	 & +151-58 & 338 & +5-177\\
105969	 & 862	 & +34-35	 & 373	 & +161-146	 &	12.4	& +1.3-1.1	&0.25	& +0.17-0.14 & 135	 & +10-8 & 126	 & +48-89 & 107	 & +178-19\\
107143	 & 1338	 & +171-73	 & 1092	 & +211-1023	 &	33.3	& +4.1-2.3	&0.14	& +0.21-0.11 & 40	 & +6-6	 & 105	 & +147-74 & 124 & +170-28\\
109095	 & 1552	 & +1278-329	 & 684	 & +891-512	 &	46.0	& +82.6-10.9	&0.40	& +0.34-0.38 & 89	 & +6-7	 & 101	 & +87-50 & 212	 & +29-160\\
110725	 & 631	 & +39-31	 & 392	 & +186-342	 &	5.08	& +8.4-1.0	&0.27	& +0.46-0.24 & 55	 & +19-15 & 93	 & +113-44 & 210 & +46-146\\
110785	 & 878	 & +65-57	 & 490	 & +125-242	 &	10.9	& +11.6-1.9	&0.29	& +0.44-0.28 & 93	 & +6-5	 & 93	 & +126-47 & 272 & +11-175\\
113136	 & 483	 & +20-19	 & 275	 & +176-232	 &	9.6	& +2.6-1.2	&0.12	& +0.25-0.15 & 131	 & +16-16 & 103	 & +72-68 & 171	 & +164-83\\
113177	 & 808	 & +26-24	 & 510	 & +160-371	 &	11.7	& +2.0-1.3	&0.18	& +0.28-0.15 & 127	 & +7-7	 & 79	 & +98-45 & 176	 & +28-167\\
113638	 & 356	 & +9-8	         & 138	 & +157-83	 &	13.1	& +6.9-1.5	&0.08	& +0.31-0.21 & 61	 & +12-13 & 104	 & +68-52 & 160	 & +176-13\\
113697	 & 1160	 & +122-71	 & 773	 & +185-150	 &	27.0	& +3.2-1.9	&0.33	& +0.20-0.13 & 139	 & +9-8	 & 92	 & +26-21 & 136	 & +13-18\\
113699	 & 680	 & +27-35	 & 181	 & +363-132	 &	9.5	& +3.1-1.3	&0.10	& +0.31-0.20 & 58	 & +13-15 & 99	 & +68-58 & 160	 & +169-26\\
117622	 & 1008	 & +161-41	 & 368	 & +564-302	 &	12.1	& +2.2-1.2	&0.25	& +0.25-0.16 & 125	 & +9-8	 & 93	 & +41-40 & 32	 & +11-11\\
118212	 & 893	 & +16-14	 & 580	 & +48-49	 &	34.1	& +2.9-2.9	&0.53	& +0.09-0.08 & 115	 & +4-4	 & 116	 & +11-12 & 112	 & +5-5\\

\enddata
\end{deluxetable}

\begin{deluxetable}{lrr}
\tabletypesize{\scriptsize}
\tablecaption{Comparison of astrometric fits with known spectroscopic orbits and 
constrained solutions. \label{known.tab}}
\tablewidth{0pt}
\startdata
\multicolumn{1}{c}{Element}  &  \multicolumn{1}{c}{HIP 34164} & \multicolumn{1}{c}{HIP 67480} \\
\multicolumn{3}{l}{Spectroscopic \dotfill}\\
$P$ [d] & $612.3\pm3.0$ & $944\pm8$ \\
$e$ & $0.273\pm0.015$ & $0.41\pm0.09$\\
$f{M}$ & $0.0826\pm0.0061$ & $0.00013\pm0.00005$\\
$\omega$ [\degr] & $248.9\pm3.7$ & $359\pm15$\\
$a_1\sin i$ [Gm] & $91.9\pm2.4$ & $14.1\pm1.9$ \\
\multicolumn{3}{l}{Astrometric, unconstrained (this work)\dotfill} \\
$P$ [d] & $643^{+91}_{-51}$ & $1167^{+153}_{-89}$\\
$e$ & $0.27^{+0.42}_{-0.31}$ & $0.38^{+0.35}_{-0.21}$\\
$i$ [\degr] & $104^{+15}_{-11}$ & $134^{+13}_{-15}$\\
$\omega$ [\degr] & $111^{+65}_{-53}$ & $111^{+45}_{-42}$\\
$a_0$ [mas] & $9.3^{+19.8}_{-2.3}$ & $9.6^{+6.0}_{-1.8}$\\
\multicolumn{3}{l}{Astrometric, constrained \citep{jan}\dotfill} \\
$i$ [\degr] & $107.4\pm8.5$ & $174.0\pm0.5$\\
$a_0$ [mas] & $8.77\pm0.96$ & $7.3\pm0.9$\\
\enddata
\tablecomments{Spectroscopic data from \citep{lat} and \citep{gri}}
\end{deluxetable}

\end{document}